# Validation and Optimization of Multi-Organ Segmentation on Clinical Imaging Archives

Olivia Tang[*a], Yuchen Xu[*a], Yucheng Tang[**a], Ho Hin Lee[a], Yunqiang Chen[b], Dashan Gao[b], Shizhong Han[b], Riqiang Gao[a], Michael R. Savona[c], Richard G. Abramson[d], Yuankai Huo[a], Bennett A. Landman[a,d]

[a]Department of Electrical Engineering and Computer Science, Vanderbilt University, Nashville, TN, USA 37212;
[b]12 Sigma Technologies, San Diego, CA, USA 92130;
[c]Hematology and Oncology, Vanderbilt University Medical Center, Nashville, TN, USA 37235
[d]Radiology, Vanderbilt University Medical Center, Nashville, TN, USA 37235

(*These authors contributed equally to this work **Corresponding author: yucheng.tang@vanderbilt.edu)

**ABSTRACT**

Segmentation of abdominal computed tomography (CT) provides spatial context, morphological properties, and a framework for tissue-specific radiomics to guide quantitative Radiological assessment. A 2015 MICCAI challenge spurred substantial innovation in multi-organ abdominal CT segmentation with both traditional and deep learning methods. Recent innovations in deep methods have driven performance toward levels for which clinical translation is appealing. However, continued cross-validation on open datasets presents the risk of indirect knowledge contamination and could result in circular reasoning. Moreover, "real world" segmentations can be challenging due to the wide variability of abdomen physiology within patients. Herein, we perform two data retrievals to capture clinically acquired deidentified abdominal CT cohorts with respect to a recently published variation on 3D U-Net (baseline algorithm). First, we retrieved 2004 deidentified studies on 476 patients with diagnosis codes involving spleen abnormalities (cohort A). Second, we retrieved 4313 deidentified studies on 1754 patients without diagnosis codes involving spleen abnormalities (cohort B). We perform prospective evaluation of the existing algorithm on both cohorts, yielding 13% and 8% failure rate, respectively. Then, we identified 51 subjects in cohort A with segmentation failures and manually corrected the liver and gallbladder labels. We re-trained the model adding the manual labels, resulting in performance improvement of 9% and 6% failure rate for the A and B cohorts, respectively. In summary, the performance of the baseline on the prospective cohorts was similar to that on previously published datasets. Moreover, adding data from the first cohort substantively improved performance when evaluated on the second withheld validation cohort.

## 1. INTRODUCTION

Abdominal multi-organ segmentation of computed tomography (CT) image aids in a variety of clinical tasks, including diagnosis, evaluation, and treatment delivery [1]. As such, segmentation of abdominal organ scans has been the subject of extensive research. While segmentations on carefully selected, well-controlled research cohorts are successful using existing models, the generalizability of these algorithms to clinical datasets remains a question (see Figure 1) [2]. The shape and distribution of abdominal organs can vary significantly between patients, as well as within a person over time depending on patient position and movement at the time the scan is taken [2]. Additionally, knowledge contamination due to continued cross-validation disrupts the efficacy of models trained on controlled research cohorts, which may cause the resultant algorithms to fail when tested on clinical imaging archives.

Current research on multi-organ abdominal segmentation has focused on the generation of 3D volumetric segmentations through integration of convolutional neural networks with manual annotation, with a 2015 MICCAI challenge leading to the development of an application of the 3D U-Net deep learning architecture [4-5]. Several algorithms have been designed to perform automatic multi-organ segmentation using a variety of techniques, including multi-atlas segmentation (MAS) [1-3]. Many attempts at site-specific customization have also spurred movement toward efficient segmentation of CT images [6-8]. Historically, these methods have proved compelling in controlled testing cases but have yet to be applied effectively to large-scale, clinically acquired CT scan data.

In this paper, we explore prevention of knowledge contamination through prospective validation on two large-scale testing cohorts. Additionally, we develop a new model trained on manually labeled failure scans of the liver and gallbladder, as it is often difficult to distinguish between these two organs through existing automatic segmentation techniques. We evaluate the performance improvement of the optimized algorithm as compared to the baseline to determine the generalizability of deep learning algorithms to a clinical dataset.

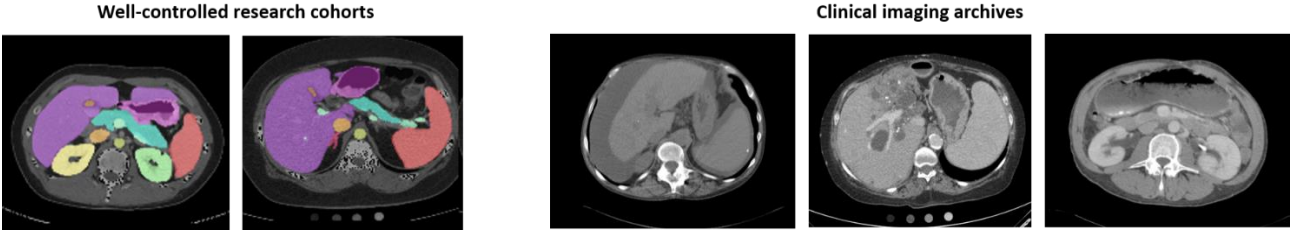

Figure 1. Multi-organ segmentation of research cohorts compared to clinical imaging archives. Scans in research cohorts are well-controlled and have been shown to allow for effective segmentation, whereas clinical scans display wide variability. Clinical scans display tumors, liver disease, interventions, etc., which make automatic multi-organ segmentation challenging.

## 2. METHOD

In this paper, we present an algorithm for optimizing multi-organ segmentation to generalize to clinical images. Failure scans from the baseline algorithm are manually labelled and introduced into the training dataset under a 3D U-Net architecture and evaluated under a large-scale testing cohort of clinical images, as seen in Figure 2.

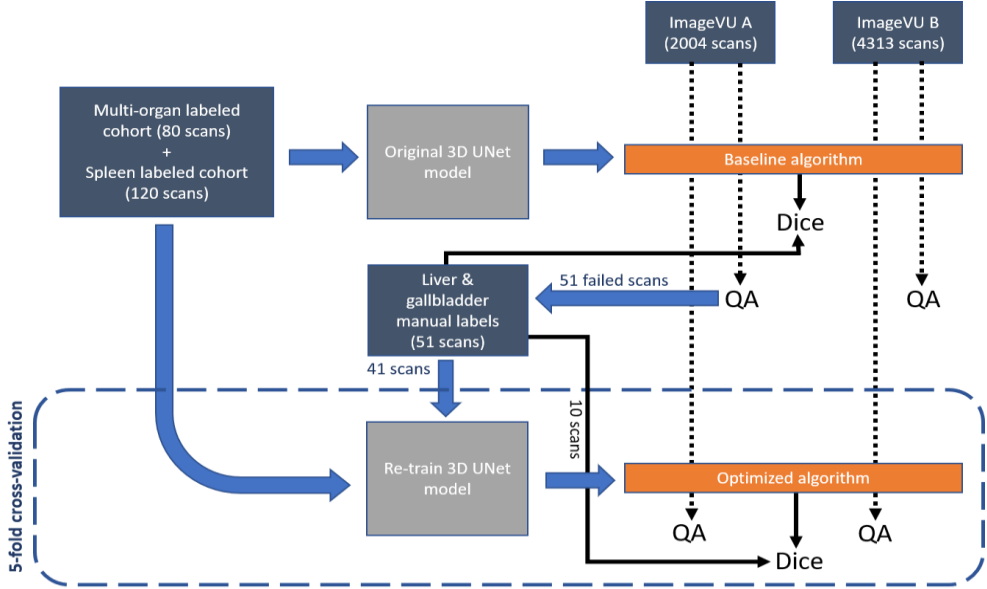

Figure 2. Training and testing of baseline algorithm on clinical imaging cohorts A and B with manual QA, leading to manual labeling of failure scans. Re-training with manual labels added to produce optimized algorithm, with manual QA performed again to compare results from baseline to optimized algorithm. Blue arrows represent processes, while black arrows lead to final outputs.

## 2.1 Baseline Algorithm

The baseline model is trained using an 80-scan cohort with multi-organ segmentation and a 120-scan cohort with spleen segmentation, forming a 200-scan training dataset. All 200 scans are trained under the existing 3D U-Net model, producing the baseline algorithm for multi-organ segmentation. In testing the baseline algorithm, we first retrieve 2004 studies on 476 patients with diagnosis codes involving spleen abnormalities (cohort A). Next, we retrieve 4313 studies on 1754 patients without diagnosis codes involving spleen abnormalities (cohort B). All studies were retrieved in deidentified form under IRB approval. Prospective evaluation of the baseline algorithm is performed on both cohorts with manual quality assurance (QA) on the generated segmentations. The segmentations are rated using a multi-level quality grading system of 0, 1, or 2 (Excellent, Usable, or Global fail). A grading value of 0 is defined as the segmentations where the algorithm visually correctly predicts almost all organs in the abdomen. A grading value of 1 is defined as segmentation where the algorithm visually correctly predicts most major organs, such as the liver and spleen. A grading value of 2 is defined as segmentations with major inconsistencies visible within organs. We observe performance of 13% and 8% global failure rate for both cohorts, respectively.

## 2.2 Manual Annotation

Taking 51 global failure scans from cohort A (splenomegaly cohort), we manually annotate the liver and gallbladder on the original images. A window and level of 35 and 190 Hounsfield units (HU) are used, respectively. Annotations are made on the axial slices of the scans while also consulting the sagittal and coronal views. Livers in many of the failed scans show signs of cirrhosis, displaying a significant amount of surrounding fluid, which is not included in the labels. Transjugular intrahepatic portosystemic shunts (TIPS) embedded in the hepatic are also present in many scans. Since the hepatic vein runs through the liver, the TIPS are included in the liver labels to maintain continuity of the organ. The hepatic and portal veins are also included in the manual labels given they are within the main contours of the liver, though any fluid within the organ is excluded. These annotation guidelines are followed consistently for all slices of all images.

When annotating the gallbladder, the manual labels are checked against the liver labels to ensure no overlap between the manual segmentations of the two organs. The gallbladder is present in only 24 out of 51 scans, having likely been removed from the remaining patients in scans where it does not appear. The consistency of the hand label process is checked via a second party manually labelling 5 randomly selected scans from the 51 images and calculating the Dice loss between the two hand tracings. This manual protocol yielded an average Dice similarity metric of 95% between the two manual labels for both organs.

## 2.3 Preprocessing:

The manual labels and corresponding images are preprocessed in three steps. First, the original scans are saved under a soft tissue window. Second, the original images, soft-tissue windowed images, and labels are normalized and resampled using spline interpolation. Lastly, all images, soft images, and segmentation labels are cropped and resampled from dimension 512x512x52 to 168x168x64.

## 2.4 Optimized Algorithm

To evaluate feasible performance improvement, the manual labels of the failure scans are added to the training dataset and the existing model is re-trained under a 3D U-Net architecture with 5-fold cross-validation. The new model is trained via transfer learning using the existing model trained on the original 200 scans (120-scan splenomegaly cohort and 80-scan multi-organ cohort). 41 hand-labelled liver and gallbladder scans are added to the training dataset for a total of 241 training scans. The 10 remaining manually labelled scans are placed in the testing cohort during each fold of cross-validation and used to calculate the average Dice value of the optimized algorithm. The model is trained for 150 epochs for each fold, stopping once the Dice value improves <0.001 across 4 consecutive epochs. The optimized algorithm is then tested using the same large-scale clinical imaging cohorts A and B. Prospective validation of the optimized algorithm is performed on both cohorts with manual QA using the same multi-level quality grading scale (0, 1, 2) to assess the failure rate of the new algorithm. The 51 manually labelled scans are also tested against the baseline algorithm for 150 epochs, and the Dice score for each image is averaged across epochs to calculate the baseline algorithm Dice value.

## 3. RESULTS

Figures 3 and 4 compare the Dice values of the liver and gallbladder in the baseline and optimized algorithms. The Dice values presented for the baseline algorithm in Figure 3 are averaged across all epochs for the 51 manually labeled scans. The Dice values presented for the optimized algorithm in Figure 3 are averages across all epochs for the 10 scans placed in each testing cohort of 5-fold cross-validation. Based on a paired t-test, the Dice values of the optimized algorithm are significantly superior to those of the baseline algorithm for both the liver and gallbladder (p<0.001). Figure 5 compares the failure rate of the baseline and optimized algorithms amongst testing cohorts A and B. Based on a Chi-squared test, the failure rate of the optimized algorithm is significantly less than that of the baseline algorithm (p<0.001). Figures 6 and 7 display the qualitative results.

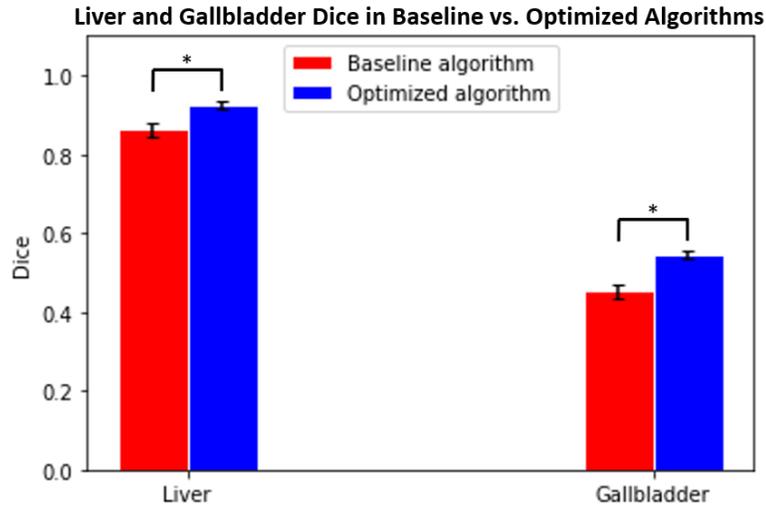

Figure 3. Average Dice values of 51 manually labeled scans calculated across all epochs in liver and gallbladder between baseline and optimized algorithms (p<0.001 using paired t-test). Error bars indicate standard deviation.

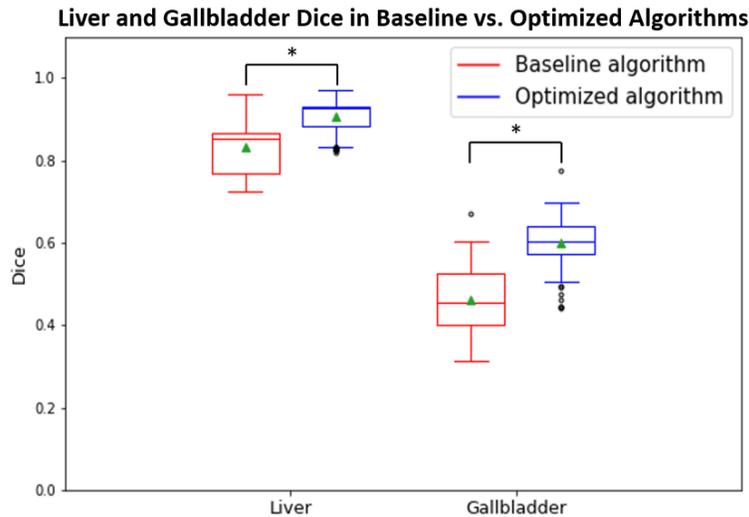

Figure 4. Dice values of 51 manually labelled scans calculated at the final epoch in liver and gallbladder between baseline and optimized algorithms (p<0.001 using paired t-test).

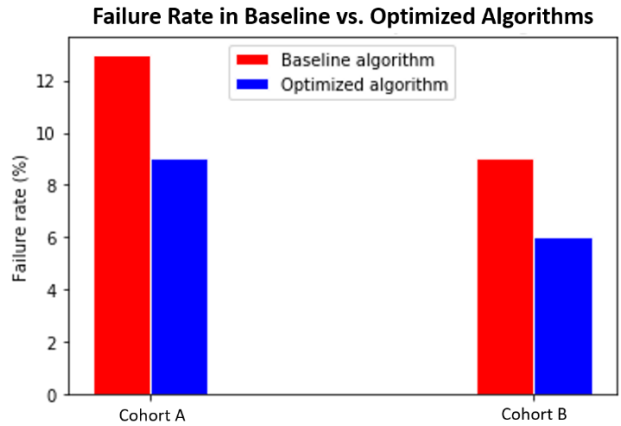

Figure 5. Failure rates in clinical imaging archive cohorts A and B are significantly reduced between baseline and optimized algorithms ($p<0.001$ using Chi-squared test).

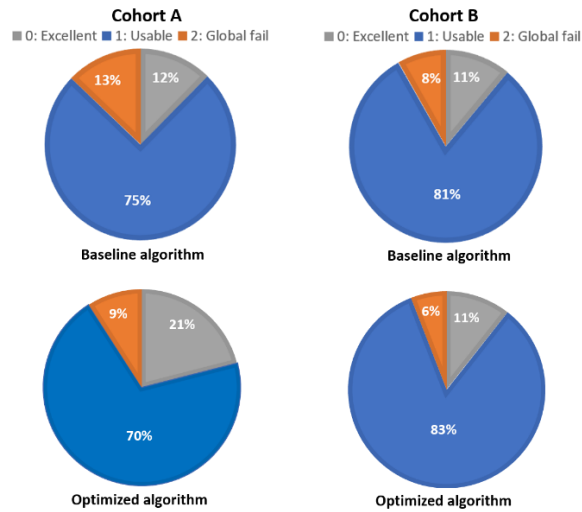

Figure 6. Cohorts A and B quality ratio breakdowns, displaying percentage of 0's, 1's, and 2's in generated segmentations from both testing cohorts between baseline and optimized algorithms.

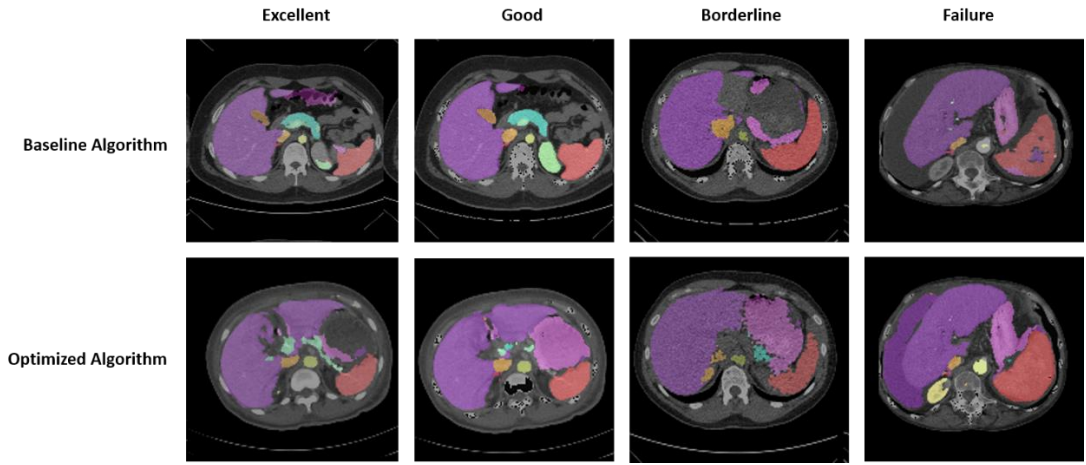

Figure 7. Four levels of multi-organ segmentation quality using the same patient scans. Scans from each level either remained generally the same in quality or improved significantly in quality between the baseline and optimized algorithms.

## 4. CONCLUSION

The method proposed in this paper generates an optimized algorithm that generalizes from controlled research datasets to clinical imaging datasets when integrated with site-specific customization. As seen in Figures 3 and 4, both the mean and the median Dice values of the 51 manually identified failure scans increased from the baseline to the optimized algorithm on the liver and gallbladder. Additionally, the failure rate decreased when testing the optimized algorithm on clinical imaging cohorts A and B, shown in Figures 5 and 6. This improvement is seen qualitatively in Figure 7. The generalizability of the optimized algorithm supports the transition of automated multi-organ segmentation from research to clinical settings. While the optimized algorithm yields substantive performance improvement over the baseline on the withheld dataset under prospective evaluation, further exploration may be undertaken to refine the algorithm in regard to reducing outliers.

## 5. ACKNOWLEDGEMENTS

This research is supported by Vanderbilt-12Sigma Research Grant, NSF CAREER 1452485, NIH grants 1R01EB017230 (Landman).This study was in part using the resources of the Advanced Computing Center for Research and Education (ACCRE) at Vanderbilt University, Nashville, TN. We gratefully acknowledge the support of NVIDIA Corporation with the donation of the Titan X Pascal GPU used for this research. The identified datasets used for the analysis described were obtained from the Research Derivative (RD), database of clinical and related data. The imaging dataset(s) used for the analysis described were obtained from ImageVU, a research repository of medical imaging data and image-related metadata. ImageVU and RD are supported by the VICTR CTSA award (ULTR000445 from NCATS/NIH) and Vanderbilt University Medical Center institutional funding. ImageVU pilot work was also funded by PCORI (contract CDRN-1306-04869).